# New Families of Carbon Nanotubes

Vitor R. Coluci, Scheila B. Furtado, Sergio B. Legoas, and Douglas S. Galvão*
*Instituto de Física, Universidade Estadual de Campinas, 13083-970 Campinas, São Paulo, Brazil*

Ray H. Baughman
*NanoTech Institute and Department of Chemistry,
University of Texas at Dallas, P.O. 830688, Richardson, Texas*

Fundamentally new families of carbon single walled nanotubes are proposed. These nanotubes, called graphynes, result from the elongation of covalent interconnections of graphite-based nanotubes by the introduction of *yne* groups. Similarly to ordinary nanotubes, arm-chair, zig-zag, and chiral graphyne nanotubes are possible. Electronic properties, predicted using tight-binding and *ab initio* density functional methods, show a rich variety of metallic and semiconducting behaviors.

* Corresponding author: galvao@ifi.unicamp.br  FAX: +551937885376

The early report of carbon nanotubes by Iijima [1] generated an enormous amount of research activity. New and exciting phenomena have been observed [2], including field emission [3], quantum conductance [4], superconductivity [5], and higher thermal conductivity than diamond [6]. Depending upon structure, the nanotubes are either metallic or insulating, which is a feature that has been intensively investigated and exploited in prototype devices [2].

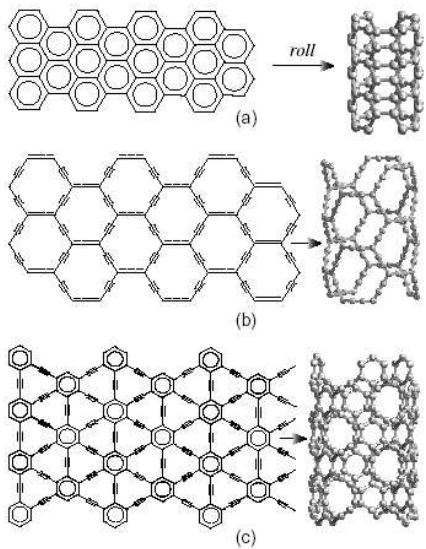

**Figure 1** – The structural relationships between carbon sheets and single wall carbon nanotubes for sheets of (a) graphene (a graphite sheet), (b) α-graphyne, and (c) graphyne. Depending upon the axis used for rolling the carbon sheet to make a seamless cylinder, the nanotube is armchair (a and b, right), zig-zag (c, right), or chiral. See text for discussions.

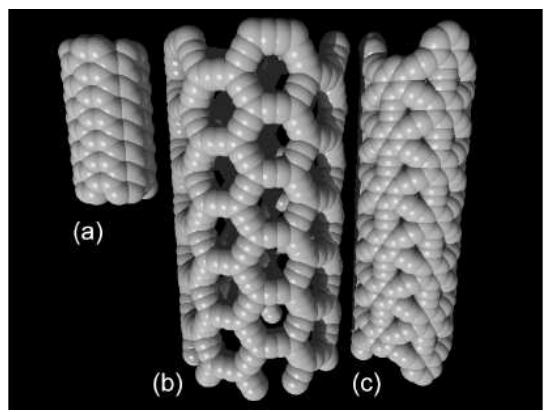

**Figure 2** – Three dimensional view of (a) conventional carbon nanotube (CNT); (b) α-GNT; and; (c) nanotube formed on graphyne sheet (GNT) (Fig. 1).The increase in nanotube sidewall porosity in going from CNT to GNT and α-GNT is evident. The same van der Waals radius values were used for all structures.

Alternative structures containing heteroatoms (N, B, etc.) in the nanotubes, as well as various carbon free nanofibers, have been recently synthesized [7]. While previous work have focused on graphitic nanotubes, we believe that other types of pure carbon nanotubes are feasible using different accessible hybridization states of carbon. One possibility that has been overlooked in the literature is to use graphyne sheets as structural motif for carbon nanotubes. Graphyne is an allotropic form of carbon proposed by Baughman, Eckhardt, and Kertesz [8] in 1987, which has recently become the focus of new investigations [9, 10]. Graphyne [Fig. 1(c)] is the name for the lowest energy member of a family of carbon phases consisting of planar molecular sheets containing only *sp* and $sp^2$ carbon atoms. The presence of acetylenic groups in these structures introduces a rich variety of optical and electronic properties that are quite different from ordinary carbon nanotubes. Acetylenes continue to provide a seemingly inexhaustible reservoir of novel materials, with extended conjugated systems [11].

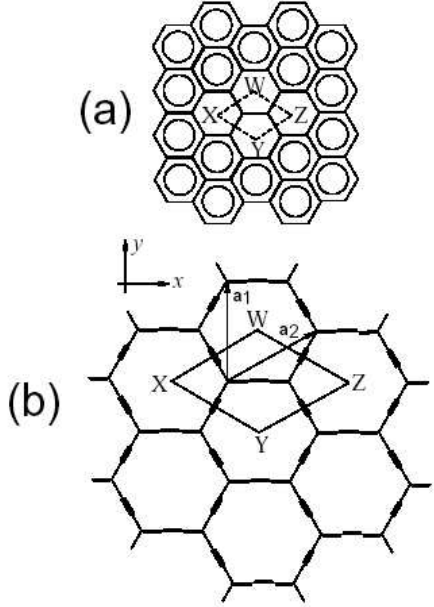

**Figure 3** – Schematic chemical structure and unit cell representation for (a) graphene, and; (b) α-graphyne [8] sheets. The lattice vectors are given by $a_1 = a y$ and $a_2 = 0.5\ a\ (\sqrt{3} x + y)$. The construction of the unit cell XYZW in (b) is straightforwardly obtained from iths equivalent in (a).

In analogy with CNTs we can imagine the graphyne nanotubes (GNTs) formation through rolling up graphyne sheets to form seamless cylinders (Figs. 1 and 2). As there are many members in the graphyne family [8] new families of pure carbon nanotubes can be generated with different electronic and structural characteristics. Just as a sheet of graphite can be rolled to form different types of nanotubes, armchair, zig-zag, or chiral GNTs are possible and the usual *(n,m)* nomemclature can be preserved. Because of space constraints, we here describe only the results for zig-zag (n,0) and armchair (n,n) α-graphyne nanotubes (α-GNTs) [Fig. 1(b)]. This structure choice is based on the fact that α-graphyne is the most analogous to graphene. Both have hexagonal unit cell (Fig. 3) and the highest possible space group (*p6m*) which facilitates the comparison between α-GNTs and CNTs. Also α-graphyne has the smallest number of atoms (only 8) in the unit cell for the graphyne families [8].

The geometrical data used for the carbon bonds were: C-C≡C-C 1.4, 1.2, 1.4 and C=C=C= 1.34, 1.32, and 1.34 Å, respectively. Bond angles of 120 and 180 degrees were used for the $sp^2$ and $sp$ carbon atoms. The corresponding lattice parameter is $a = 4 \times \sqrt{3}$ Å.

In order to directly compare results for CNTs and GNTs, we initially use the methodology of Saito et al. [2,12] and Wallace [13]. Using the 'slicing' process [2], results for α-GNTs are derived from the band structure that we predict for a planar sheet of α-graphyne.

Following the Wallace [13] methodology for tight-binding calculations, each carbon atom is described by one $2p_z$ orbital with first-neighbors interactions. The molecular orbital energy and hopping integral values were calculated in the Extended Hückel framework [14], using the parameterization proposed by Clementi and Raimondi [15]. The overlap integrals were calculated using the Mülliken *et al.* approach.

In Fig. 4 we show the TB band structure results for 2D α-graphyne. We can see that the Brillouin zone (BZ) has a different symmetry than for graphene. Also the band crossing (which characterizes the metallic behavior) occurs at different k-point ($\mathbf{k}^*$)

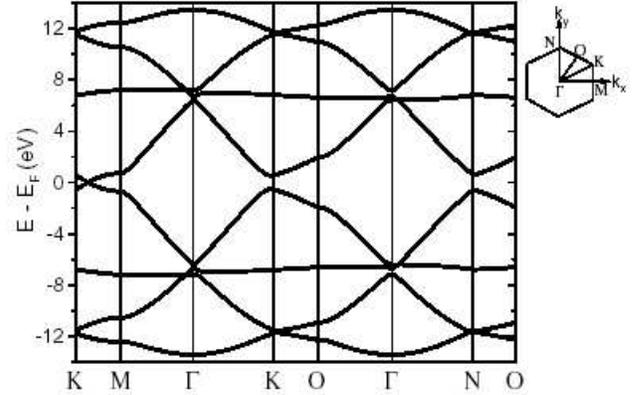

**Figure 4** – Band structure of α-graphyne sheet obtained from TB calculations. The Brillouin zone is also shown. $E_F$ is the Fermi energy. A band crossing characterizing metallic behavior can be seen in the K-M direction. The k-point of the intersection is $\mathbf{k}^* = (2/\sqrt{3}\ \pi\ a)\ \mathbf{x} + \alpha\ (\pi/a)\ \mathbf{y}$, where $\alpha \approx 0.48489$ rather than 2/3 for the graphene case.

Through the 'slicing' process [2] the band structure of α-GNTs is obtained. Typical results for (n,m) α-GNTs are displayed in Fig. 5. Similarly to CNTs, all arm-chair α-GNTs present metallic behavior. Zig-zag CNTs, which have the parameter (n,0), are close to metallic when n is multiple of three, otherwise they are semiconductors. The equivalent rule for α-GNTs is obtained from the condition (see Fig. 4):

$$k^*_y = (2\pi\ l_n) / (a\ n),\ l_n = 1, 2, 3, ..., n \quad (1)$$

which leads to:

$$n \approx 4.125\ l,\ l = 1, 2, 3, ... \quad (2)$$

i.e., a fractional number rule for the metallic behavior of the zig-zag α-GNTs. This behavior has not been previously observed for any carbon nanotube type (either pure carbon or containing heteroatoms). These results can be better visualized in Fig. 6. The expected gap decrease with increasing GNT diameter is evident, as is an oscillatory behavior of bandgap with increasing values of n. Hence, all the zig-zag α-GNTs have rather large bandgaps until the nanotube diameter becomes very large. The smallest zig-zag GNT having an essentially zero bandgap is (33,0) GNT. The porosity of all the graphyne-based nanotubes is also an interesting aspect, since this porosity should facilitate materials transport through the nanotube sidewalls, which could be important for materials storage and electrochemical charging process. In fact, selected dopants could even reside in the pore volume of the nanotube walls.

Although the TB approach is quite simple, it reliably predicts the main electronic features for α-graphyne. For comparison with the TB calculation results of Fig. 4, Fig. 7 shows that band structure that we calculate for a planar 2-D-α-graphyne sheet using an *ab initio* density functional method. The SIESTA code [17] and the above described bond lengths and bond angles were used. We also used the generalized gradient approximation (GGA) with a Perdew-Burke-Ernzerhof exchange-correlation functional [18]. The substantial agreement between the key electronic features predicted by the TB and *ab initio* method validates the TB for this application.

The stability of forms of graphyne and methods of synthesis are key remaining issues. Among the graphyne families α-graphyne has probably the highest heat of formation in relation to graphene. We estimate it to be almost 1 eV / C higher in energy from a sheet of graphite making them in principle very difficult to synthesize. Besides their very unusual electronic properties they might even be explosive.

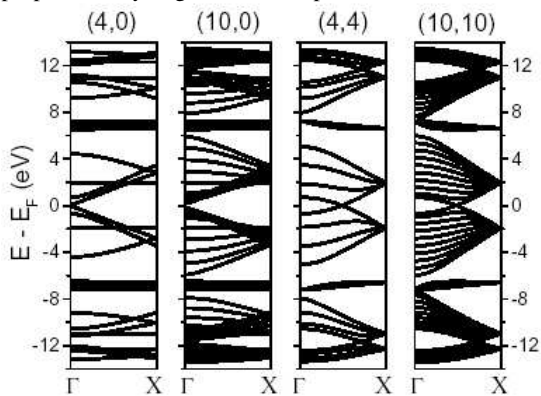

**Figure 5** – Band structure of zig-zag, and armchair α-GNTs. The bandgaps are 0.09 eV and 0.44 eV for (4,0) and (10,0) GNTs, respectively. The armchair α-GNTs has no bandgap.

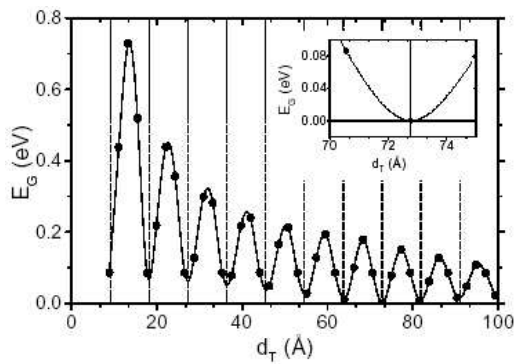

**Figure 6** – Dependence of the gap energy ($E_G$) of zig-zag α-GNTs with the tube diameter ($d_T$). The diameters correspond to tubes with *n* in the range $4 \leq n \leq 45$. Vertical lines indicate the diameters where a zero gap is expected. The inset graph shows the first allowed α-GNTs with practically zero gap which corresponds to the (33,0) tube.

A more attractive synthetic target would be graphynes [Fig. 1(c)] since they have both lower energy and much higher expected stability than does α-graphyne. Graphyne is predicted to be about 0.5 eV / C lower in energy than α-graphyne. The molecules shown in Fig. 8 have been synthesized, as well as more complex molecules that contain the carbon bonding arrangement of graphyne [19, 20]. Since conversion of either planar graphyne or a graphyne nanotube to graphite and a ordinary single walled carbon nanotube, respectively, would require the rupture of at least one carbon for every six carbons, high stability is suggested. High stability is also found for known molecules that contain fragments of graphyne structure. Efforts to synthesize both planar graphynes and GNTs are now in progress, using routes involving the topochemical polymerization of precursor molecules.

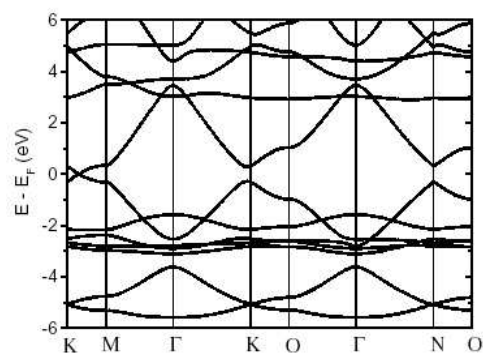

**Figure 7** – The band structure predicted for a sheet of α-graphyne using *ab initio* density functional calculations. The main features observed here agree with those shown in Fig. 4 for the tight-binding calculations.

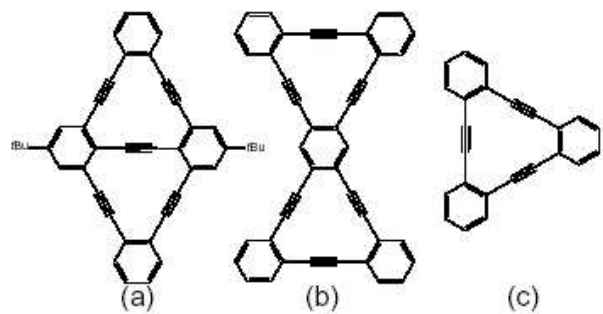

**Figure 8** – Available molecules that contain the carbon bonding pattern found in graphyne.

In summary, the proposed new families of single walled carbon nanotubes show even richer variation in electronic properties than do ordinary single-wall nanotubes. Like for usual carbon nanotubes, the armchair α-GNTs are metallic. In contrast, the zig-zag α-GNTs have electronic properties that are represented by an unusual fractional number rule. The bandgap oscillates with increasing nanotube diameter, and converges to zero more slowly than for ordinary single walled carbon nanotubes. Since planar graphyne is a relavatively large bandgap semicondutor we expect that all GNTs will be semiconductors – independent of the wrapping angle. However, depending upon the structure, both metallic or semiconducting behavior is expected for doped GNTs. The holes in the graphyne nanotube shells will enable unprecedented shell doping, as well as rapid materials transport through the nanotube sidewalls. Reflecting interest in these unusual properties, we are now trying to make GNTs and expect the present results can stimulate further studies on graphyne structures.

The authors wish to thank the Brazilian Agencies CNPq, FINEP, and FAPESP for financial support, and the CENAP-SP for computational assistance.